\documentclass[fleqn,usenatbib]{mnras}
\usepackage{amsbsy}
\usepackage{amssymb}
\usepackage{amsmath}
\usepackage{graphicx}
\usepackage{xcolor}
\usepackage[caption = false]{subfig}
\usepackage [english]{babel}
\usepackage [autostyle, english = american]{csquotes}
\usepackage{natbib}
\MakeOuterQuote{"}

\title [Modulation of pulse profile] 
{Modulation of pulse profile as a signal for phase transitions in a pulsar core}
\author [Bagchi, Layek, Sarkar and Srivastava]
{Partha Bagchi $^1$ \thanks{E-mail: parphy85@gmail.com, parphy@tsinghua.edu.cn}, 
Biswanath Layek $^2$ \thanks{E-mail: layek@pilani.bits-pilani.ac.in},  
Anjishnu Sarkar $^3$ \thanks{E-mail: anjishnu@lnmiit.ac.in} \& 
Ajit M. Srivastava $^4$ \thanks{E-mail: ajit@iopb.res.in}] \\
$^1$ Physics Department, Tsinghua University, Beijing-100084, China.\\
$^2$ Department of Physics, Birla Institute of Technology and Science, Pilani-333031, India. \\
$^3$ Physics Department, The LNM Institute of Information Technology, Jaipur-302031, India. \\
$^4$ Institute of Physics, Sachivalaya Marg, Bhubaneswar-751005, India.} 
\date{}
\pubyear{}
\begin{document}
\label{firstpage}
\pagerange{\pageref{firstpage}--\pageref{lastpage}}
\maketitle

\begin{abstract}

 We calculate  detailed modification of pulses from a pulsar
arising from the effects of phase transition induced density fluctuations 
on the pulsar moment of inertia. We represent general statistical 
density fluctuations using a simple model where the initial moment of 
inertia tensor of the pulsar (taken to be diagonal here) is assumed 
to get  random additional contributions for each of its component which are 
taken to be Gaussian distributed with certain width characterized by
the strength of density fluctuations $\epsilon$. Using sample values of 
$\epsilon$, (and the pulsar deformation parameter $\eta$) 
we numerically calculate detailed pulse modifications by 
solving Euler's equations for the rotational dynamics of the pulsar.
We also give analytical estimates which can be used for arbitrary values of
$\epsilon$ and $\eta$.  We show that there are very specific
patterns in the perturbed pulses which are observable in terms of
modulations of pulses over large time periods. In view of the fact that 
density fluctuations fade away eventually leading to a uniform phase in the 
interior of pulsar, the off-diagonal components of MI tensor also vanish 
eventually. Thus, the modification of pulses due to induced
wobbling (from the off-diagonal MI components) will also die away eventually.
This  allows one to distinguish these transient pulse modulations from the
effects of any wobbling originally present. Further, the decay of these
modulations in time directly relates to relaxation of density fluctuations
in the pulsar giving valuable information about the nature of phase transition
occurring inside the pulsar.

\end{abstract}

%\begin{keywords} pulsar, pulse profile, precession, phase transition, density fluctuations.
%\end{keywords}

%%%%%%%%%%%%%%%%%%%%%%%%%%%%%%%%%%%%%%%%%%%%%%%%%%%%%%%%%%%%%%%%%%%%%%%%%
\section{Introduction}
\label{section:sec1}

 For last several decades, QCD phase diagram is being intensely
investigated both, on theoretical front as well as on experimental
front. The high temperature and low baryon chemical potential regime of 
strongly interacting matter has been thoroughly studied at RHIC and LHC which
have provided compelling evidence for the formation of quark-gluon
plasma (QGP). This regime is very interesting as it closely resembles 
the state of matter during first few microseconds of the early stages 
of the Universe. There is compelling theoretical evidence that high baryon
density regime of QCD provides an extremely rich landscape. Starting with
the possibility of transition to high baryon density QGP, there are exotic
phases of QCD expected at much larger baryon densities, such as color 
flavor locked (CFL) phase, 2SC phase, quarkyonic phase, crystalline color 
superconductor phase etc. \citep{raja}. However, most of these phases are 
expected to occur at very high values of  baryon chemical potential which 
are difficult to achieve in laboratory experiments. Focused experimental 
efforts are underway/planned, such as the beam energy scan (BES) program of 
RHIC, CBM at FAIR, and NICA. Though, the required baryon density for some 
of these exotic phases may be out of reach in these laboratory
experiments.  

 This regime of very high baryon density in the QCD phase diagram also 
relates to the cosmos, albeit in a completely different stage
of the evolution of the Universe. QCD matter in this regime is expected
to occur during present stage of the universe in the cores of compact objects,
such as a neutron star, which form at the end of life of normal stars. 
Baryon densities in the cores of these objects can reach very high values,
allowing possibility of these exotic QCD phases to occur.  

  It then becomes very important to focus efforts on the possibility of
observation of these phases in these compact astrophysical  objects. Indeed,
many signals have been proposed in literature to probe these 
phases \citep{heisel}.
Some of us had earlier proposed a technique \citep{bagchi15} to probe the 
possibility of phase transitions occurring in the interior of a 
pulsar (which is a rotating neutron star) utilizing the fact that measurements 
of pulse timings of pulsar signals have reached extraordinary precision, to 
the level of one part in 10$^{15}$.  Basic physics of that approach is based 
on the fact that any phase transition necessarily leads to density 
fluctuations. These density fluctuations will be statistical in nature, and
will be transient, eventually subsiding and leading to the uniform new phase of
the matter. Such density fluctuations arising during a phase transition
occurring inside the core of a neutron star will lead to transient changes in 
its moment of inertia (MI) tensor. This will directly affect its rotation,
and hence the pulsar timings. With extremely accurate measurements of pulsar 
timings, very minute changes of moment of inertia of star may be observable,
providing a sensitive probe for phase transitions in these objects. 

 As emphasized in ref. \citep{bagchi15}, there are two main aspects of this 
proposal which can make this technique a powerful probe of phase transitions 
inside neutron stars. First is that the resulting density fluctuations being 
statistical in nature, every component of MI tensor will be affected. A 
typical neutron star has very high degree of symmetry, with extremely tiny 
difference in different MI values (of order 10$^{-4}\%$ or less) 
\citep{horowitz09, baiko18}. Density fluctuations will modify every 
component of MI tensor, and for a NS 
rotating about one of its symmetry axes, will generate non-zero off-diagonal
components of MI tensor. Consequently,  a spinning neutron star will develop 
wobble (on top of any previously present) which will lead to modulation of 
the pulse profile as the direction of the beam pointing towards earth will
now undergo additional modulation. This is a unique, falsifiable, prediction 
of this model, and helps in distinguishing such a signal of phase transition 
from the phenomenon of glitches. This is because standard explanation of 
glitches invokes de-pinning of vortex clusters in the superfluid core of
the NS. Vortices being directed along the rotation axis, primary effect
of this de-pinning will be on the spinning rate without significantly
perturbing the rotation axis itself. In contrast, density fluctuations
from phase transitions will affect diagonal as well as off-diagonal 
components of MI to same order, thus leading to same order of magnitude
effect for the pulse timing as well as the modulation of pulse profile.

 The second important feature of this technique relates to the precise
nature of density fluctuations. Specific statistical distribution of 
density fluctuations \citep{landau} arising during a phase transition, and 
the manner in which these density fluctuations decay away (leading to 
eventual uniform new phase) crucially depends on the nature of the phase transition.
For example, a first order transition proceeding via bubble nucleation
leads to specific density fluctuation pattern on the scale of bubble size
\citep{appl85,kaja86,appl87,chri96,layek}, whereas a transition happening via 
spinodal decomposition  has entirely different distribution of density 
fluctuations reaching very large length scale. Density fluctuations during 
a second order phase transition have universal nature \citep{golden}, and 
depend on the specific critical exponents associated with the phase transition. 

 An entirely different and rich source
of information is contained in these density fluctuations for spontaneous
symmetry breaking phase transitions if there are associated topological 
defects. Topological defects can occur in a variety of shapes, from point
defects (monopole), to strings, domain walls, and three dimensional 
structures called as Skyrmions. Formation of 
topological defects in symmetry breaking transitions 
is now a very mature field and there is extensive literature on this subject.
Dominant mechanism of the formation of topological defects during 
spontaneous symmetry breaking transitions is via the so called
{\it Kibble mechanism} \citep{kbl1, kbl2} which was originally proposed for
cosmic defects. Subsequently it was realized that this mechanism has
completely general applicability \citep{zurek}. Indeed, it is now used to study 
topological defect formation during any phase transition, from those
occurring in the universe, to a variety of condensed matter systems \citep{ams}, 
and in neutron star cores etc.  

These defects can be source of significant density fluctuations 
depending on the relevant energy scales. Important point is that the defect 
network resulting from a phase transition and its evolution shows universal
characteristics. For example, initial defect density depends only on the 
relevant correlation length and on the relevant symmetries (and space 
dimension). Further, the evolution of string defects and domain wall 
defects shows scaling behavior. These  universal properties of
defect network and scaling during evolution will be expected to lead to 
reasonably model independent predictions for changes in the moment of 
inertia tensor and its time dependence, hence on the effects on pulsar 
timing,  the pulse modification, and specifically, the eventual  subsequent 
relaxation to the  original state of rotation. We mention here that there will
also be an effect of the new uniform phase on pulsar rotation. Due to free 
energy difference between the two phases, pulsar rotation frequency will be 
directly affected. This aspect has been discussed in our earlier 
work \citep{bagchi15} where the possibility was discussed that such effects 
may provide an explanation for both glitches and antiglitches in a unified framework. 
In the present paper, we will not discuss the effects of net changes in the free 
energy of two phases and 
will only focus on the effects of density fluctuations. However, we mention here that 
the effects of free energy difference on pulsar timing and pulse modulations caused 
by the density fluctuations depend on different sets of parameters. It is
possible that the free energy difference may induce changes in pulsar timings which 
may not be observable at present. In contrast, the pulse modifications due to induced 
wobbling (from off-diagonal MI terms) may be within reach of observations (we will come 
back to this point at the end of section \ref{section:sec3}).

Examples of specific changes in different components of MI tensor, with crude estimates 
of the magnitudes were  given in ref. \citep{bagchi15} for phase transitions between 
different exotic  QCD phases. A particular interesting example of phase transition 
discussed in ref. \citep{bagchi15} was for the so-called nucleonic superfluid phase. 
This is not one of the exotic QCD phases alluded to in the discussion above. 
This is a rather conventional phase expected to occur inside cores of
neutron stars, and is of crucial important in explaining the phenomena of
glitches. Despite being of much lower energy scale (with the relevant
free energy density of order few tenths of MeV) compared to the exotic QCD 
phases such as the CFL phase (with relevant energy scale being
the QCD scale of 200 MeV, or higher depending on the baryon density)
even this superfluid phase transition is expected to lead to significant
changes in the MI tensor \citep{bagchi15}.  

  We follow up this proposal of \citet{bagchi15} in this paper and calculate 
specific signals resulting from these density fluctuations in terms of its
effect on the modification of the pulses of the pulsar. There being too many
possibilities for different phase transitions in the pulsar core, we
present a general study in this work where the specific details of density
fluctuations relating to particular phase transition are ignored. The only
relevant part used here is that these are random density fluctuations, and
are expected to affect each component of the MI tensor. For simplicity,
in this first study of this kind, we make further approximation and assume
that the initial moment of inertia tensor $I_{ij}^0$ gets additional 
contribution $\delta I_{ij}$ for each of its component. Initial
MI tensor is taken to be diagonal with eigenvalues $I_{33}^0 = I_0,$ and 
$I_{11}^0 = 
I_{22}^0 \equiv I_T < I_0$. Here $I_{ii}^0$ refer to $I_{xx}, I_{yy}, I_{zz}$ 
for $i = 1,2,3$ respectively.  $\delta I_{ij}$ is assumed to be Gaussian 
distributed with width $\sigma = \epsilon I_0$.  In view of the estimates in 
ref. \citep{bagchi15}, we consider two specific values of $\epsilon$ 
$10^{-8}$ to $10^{-5}$ (in order that the pulse modulations are
visible in a reasonable time scale for the numerical computations). Though,
we give analytical estimates which can be used for even lower values of
$\epsilon$ as discussed in ref. \citep{bagchi15} (it is not clear if such low 
values will lead to pulse modifications which can be currently observed). 

We will see that there are very specific patterns in the perturbed pulses 
which are observable in terms of modulations of pulses over much larger time 
periods than the basic pulse period. In view of the fact that 
density fluctuations fade away eventually leading to a uniform phase in the 
interior of pulsar, the off-diagonal components of MI tensor also vanish 
eventually. As we will discuss below, as a consequence of this, pulsar
restores its original state of rotation completely (apart from any
effects resulting from free energy changes between the two phases of
the transition as discussed above). In particular the 
modification of pulses due to induced
wobbling (from the off-diagonal MI components) will also die away eventually.
This will be crucial in distinguishing these pulse modulations from the
effects of any wobbling originally present.

 We note that in representing the effect of density fluctuations on MI
tensor in terms of Gaussian distributed random components $\delta I_{ij}$ with 
a single parameter $\epsilon$, we are missing out very useful information 
about characteristic statistics of the density fluctuations which could 
differentiate between different types of phase transitions. Thus, the present
study is meant to focus on the gross features of the pulse modification, such
as the period and amplitude of pulse modification. Next step will be to
determine the detailed modification of the MI tensor depending on specific
phase transition, and see if observations of the perturbed signal are 
capable of distinguishing between different phase transitions.

We mention that there have been
several theoretical studies on the effects of free precession of pulsars 
because of its various observational consequences. The studies of pulsar precession 
have become even more exciting and relevant 
as there seems to be evidence of free precession, as reported from the 
observation of the periodic residuals of PSR 1828-11 \citep{stairs}.
There was a theoretical proposal \citep{ira03, akgun06} that precession 
caused by triaxiality of the pulsar can be a possible cause 
for such behavior of PSR 1828-11. The precession of pulsars is also
studied to probe the internal structure of neutrons stars. In this context, 
modeling the free precession of neutrons stars and by comparing with the 
observations, the authors \citet{jones01} have made a few interesting 
conclusions regarding the crust-core coupling and on the possible role of 
superfluidity in the free precession of the crust. In our work here, 
we aim to probe the various phase transitions occurring 
inside the core of pulsars. Here, the phase transition induced density 
fluctuation is considered to be responsible for the precession of pulsars 
affecting the pulse profile.

 The paper is organized in the following manner. In section \ref{section:sec2}, 
 we present the basic formalism for calculating the effects of the modification
in the MI tensor by a random matrix resulting from density fluctuations 
on the state of rotation of the pulsar. Using Euler's equations, we calculate
the rate of change of angular velocities about the principal axes of
the pulsar (which, due to density fluctuations, differ from the original
principal axes of a symmetric spheroidal shape pulsar). Here we focus on
specific points on the surface of the pulsar which are emitting radiation
(which, again, for simplicity is taken to be on the surface of the pulsar),
and study changes in its trajectory as the pulsar rotation develops
wobbling. We then calculate the resulting perturbation in the pulsar signal 
as observed on the earth. Section \ref{section:sec3} discusses parameter choices, 
initial 
conditions, and estimates for modulation frequency etc. depending on the 
magnitudes of $\epsilon$, and $I_o,I_T$.  Section \ref{section:sec4} presents 
the algorithm for calculation of the pulse modification and presents numerical results.
Section \ref{section:sec5} presents discussion of results and various observational
aspects. We conclude in section \ref{section:sec6} with discussion of various limitations
of our procedure, and future possibilities, e.g. possibility of observing
details in the perturbed signal which can distinguish between different phase
transitions.

\section{The effects of density fluctuations on pulsar dynamics : The 
basic formalism}
\label{section:sec2}

\begin{figure}
\centering
%\vspace{-1.0cm}
\includegraphics[width=0.6\linewidth]{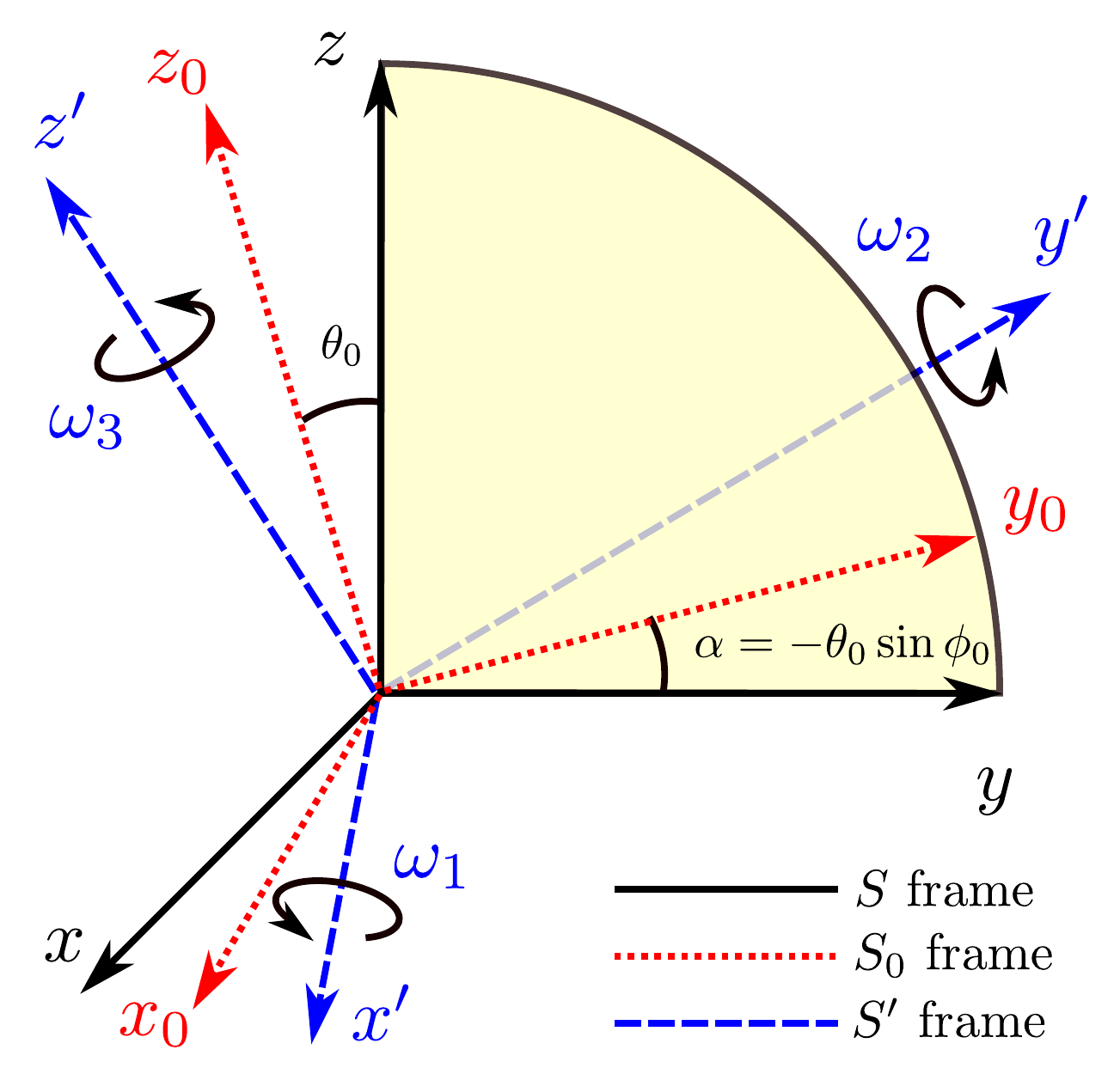}
%\vspace{-3.2in}
\caption{Before phase transition, an oblate shape pulsar is rotating about 
z-axis with frequency $\omega$ and angular momentum $L_z = L$. Black solid 
lines show the space fixed frame S for the unperturbed pulsar.
Orientations of the principal axes ($x_0$, $y_0$, $z_0$) 
immediately after the phase transition (at $t = 0$) of $S_0$ frame are shown 
with red dotted lines. The body-fixed frame $S^\prime$ at any arbitrary time 
$t$ is shown with blue dashed lines.}
\label{fig:fig1}
\end{figure}

For the study of pulsar dynamics in the presence of phase transition induced 
density fluctuations, we take the initial shape of the pulsar to be oblate 
spheroidal. The pulsar is assumed to be rotating about the symmetry $z$-axis 
with angular frequency $\omega$ and angular momentum $L_z = L$ ($L_x = 0 = 
L_y$). The unperturbed principal moment of inertia (MI) with respect to the 
body-fixed 
frame $S$ (Fig. \ref{fig:fig1}) are denoted by $I_{ij}^0$ ($i, j = 1,2,3$). 
In brief notations, the diagonal components can be written as, $I_{11}^0 \equiv 
I_1^0$, $I_{22}^0 \equiv I_2^0$ with $I^0_1=I^0_2$ and $I_{33}^0 \equiv I_3^0 
= I_0$ (with $I_0 > I^0_1, I^0_2$), and $I_{ij}^0 = 0$ for $i \ne j$. The oblateness 
of the star is parameterized 
by $\eta = (I_0 - I_1^0)/I_0$.  The value of $\eta$ depends on 
various properties of the star, such as mass, rigidity of the crust, and 
the magnetic field etc. There have been several studies 
where the authors \citep{horowitz09, baiko18} have carried out detailed 
molecular dynamical simulations to determine the values of $\eta$ by 
estimating the crustal breaking strain of neutron stars. Those works have 
put an upper limit of ellipticity as $\eta \simeq 10^{-6}$.  However, the 
results being sensitive to the modeling of the crust, there are uncertainties 
in the estimates of breaking strain. In fact, it has been suggested (based 
on the studies of a magnetar, \citep{maki14}) that the deviation from 
the sphericity of pulsar can be as high as $\sim 10^{-4}$. From observational 
perspective, there were several attempts \citep{vela06, aasi14, ligo20} to 
constrain the deformation parameter of triaxial stars through direct searches 
for gravitational waves (GWs). For example, a recent result \citep{ligo20} 
constrained the upper limit of $\eta$ for Crab and Vela pulsars at $10^{-5}$ 
and $10^{-4}$, respectively. As recorded by \citet{aasi14}, there are also 
a few pulsars with extremely high ($\eta \sim 10^{-2} - 10^{-3}$) 
ellipticities. 

Note that such constraints are valid for triaxial stars only. It is not 
possible to put such constraints for the spheroidal pulsars due to the 
absence of continuous GWs from these sources. However, within these 
observational limitations and the uncertainties in theoretical estimate, 
we will take sample values of $\eta$ in the range $10^{-3} - 10^{-2}$ for 
the initial unperturbed pulsars. This is helpful in showing modulations of
pulse profile over reasonable time duration.  The results can be 
straightforwardly extended to much smaller values of $\eta$ (which lead to 
pulse modulations over very long time durations).

As we have mentioned earlier, the phase transitions inside the core of a 
pulsar inevitably produce density fluctuations \citep{bagchi15}, and hence 
cause perturbation in MI tensor.  Importantly, the MI tensor now develops 
non-zero off-diagonal components causing the star to precess about the 
z-axis. Detailed simulations were carried out in ref. \citep{bagchi15} 
to estimate the magnitude of density fluctuations caused by various possible 
phase transitions inside the core of a pulsar. In those simulations, 
depending on the types of transitions, the fractional change in the MI 
tensor $\delta I_{ij}/I_0$ were estimated to be of order 
$\sim 10^{-14} - 10^{-6}$ (with the limitation of extrapolating the simulation
results of small lattice sizes to realistic NS size). We would like to study 
the pulsar dynamics in the presence of such perturbations and present
results for density fluctuations of this order. As we will see, the results
can be straightforwardly extended to arbitrary small values of density 
fluctuations (though, possibility of observing effects of much smaller
density fluctuations on pulse profiles may not be realistic at present stage).

Fig. \ref{fig:fig1} shows the space fixed frame $S$ (black solid lines) for the 
unperturbed pulsar (of oblate shape) which is rotating about the z-axis with 
frequency $\omega$ and angular momentum $L_z = L$. Phase transition is assumed
to occur at time t = 0 (for simplicity, we assume the transition to be
instantaneous) generating density fluctuations. With density fluctuation,
the new principal axes of the pulsar at time $t = 0$ (after the phase 
transition) are denoted by ($x_0$, $y_0$, $z_0$).  This new body fixed 
frame $S_0$ (at t = 0) is shown by red dotted lines. $S^\prime$ frame (with 
axes $x^\prime, y^\prime, z^\prime$) denotes
this body fixed frame at any arbitrary time $t > 0$  and is shown by 
blue dashed lines.

Let us denote by $I_1$, $I_2$ and $I_3$, the principal MI of the perturbed 
pulsar with respect to the body fixed frame $S^\prime$ (as in 
Fig. \ref{fig:fig1}). The frame $S^\prime$  momentarily coincides 
(at time $t$) with a space fixed frame, with respect to which the dynamical 
equations need to be written.  The angular frequency of the star about this 
frame are denoted by $\omega_1(t)$, $\omega_2 (t)$ and $\omega_3 (t)$, 
respectively. 
%%%%%%%%  FIGURE-1 %%%%%%%%%
%%%%%%%%%%%%%%%%%
The set of Euler's equations which governs the dynamics of the pulsar can 
now be written as \citep{goldstein,klepp}

\begin{align}
I_1 \dot \omega_1 - (I_2 - I_3) \omega_2 \omega_3 & =  0 \label{eq:euler1} \\
I_2 \dot \omega_2 - (I_3 - I_1) \omega_1 \omega_3 & =  0 \label{eq:euler2}\\
I_3 \dot \omega_3 - (I_1 - I_2) \omega_1 \omega_2 & =  0 \label{eq:euler3}.
\end{align}

We assumed that the angular frequency $\omega$ of the unperturbed state is 
along the z-axis.  As density perturbations are assumed to be small, the 
value of $\omega_3$ is expected to remain close to $\omega$. However, $\omega_1$ 
and $\omega_2$ now become non-zero, though much smaller compared to $\omega_3$, i.e.,  $\omega_1, 
\omega_2  << \omega_3$. Thus, within first order in $\omega_1$ and $\omega_2$, 
we get from Eq. (\ref{eq:euler3})

\begin{equation}\label{eq:omega3}
I_3 \dot \omega_3  = 0 ~; \text{i.e.,}~~ \omega_3 = \text{constant}.
\end{equation}

Here, we have neglected the higher order term $\omega_1 \omega_2$ in 
Eq. (\ref{eq:euler3}). The dynamics of $\omega_1$ can now be 
obtained by using Eq. (\ref{eq:euler1}) and Eq. (\ref{eq:euler2}) as

\begin{equation}\label{eq:diff}
\ddot \omega_1 + \Omega^2~ \omega_1 = 0 .
\end{equation}

Where, $\Omega = \omega_3 [(I_3 - I_1)(I_3 - I_2)/
(I_1 I_2)]^{1/2}$ is the precession frequency of the pulsars caused by 
the perturbations. We assume that phase transition induced density fluctuations
are sufficiently small so that $I_3 > I_1, I_2$ condition remains valid
even after the phase transition. Thus, $\Omega$ will be real. With this,
the solution of Eq. (\ref{eq:diff}) becomes oscillatory, 

\begin{equation}\label{eq:omega1}
\omega_1(t) = A~\cos (\Omega~t) + B~\sin (\Omega~t).
\end{equation}

$A$, $B$ are two constants which can be determined from the initial 
conditions. The solution of $\omega_2$ can be obtained by simply finding 
$\dot \omega_1$ from Eq. (\ref{eq:omega1}), and substituting it in 
Eq. (\ref{eq:euler1}). The resulting solution is given by

\begin{equation}\label{eq:omega2}
\omega_2 (t) = k [ A \sin (\Omega~t) - B \cos (\Omega~t)].
\end{equation}

Where the overall factor $k$ is given as 
$k = [I_1(I_3 - I_1)/(I_2(I_3 - I_2))]^{1/2}$. 
We have set the initial time $t = 0$ as the time of 
completion of the phase transition, which is assumed to be the 
onset of precession of the pulsar as well. If $\omega_1^0$ and  $\omega_2^0$ 
denote the respective angular velocities at $t = 0$, the set of solutions 
(\ref{eq:omega1}), (\ref{eq:omega2}) can then be rewritten as (using
Eqs. (\ref{eq:euler1}), (\ref{eq:euler2})),

\begin{align}
\omega_1 (t) & =   \omega_1^0 \cos (\Omega~t) - \frac{\omega_2^0}{k}  
\sin (\Omega~t) \label{eq:omega10} \\
\omega_2 (t) & =  k \omega_1^0 \sin (\Omega~t) + \omega_2^0  
\cos (\Omega~t).\label{eq:omega20}
\end{align}

Note that the above set of equations still has two arbitrary constants 
$\omega_1^0$ and $\omega_2^0$ to be fixed from the initial conditions. In 
the next section, we will discuss the procedure of fixing these quantities 
from the given initial conditions. Along with this, we will also discuss 
our choice of parameters, and present an estimate of precession 
frequency $\Omega$. More detailed numerical procedures for obtaining the 
effects on pulse profiles will be discussed in the subsequent section.
 
\section{Initial conditions, choice of parameters, and estimates for 
modulation frequency}
\label{section:sec3}

\subsection{Initial conditions} 
The values of $\omega_1^0$ and $\omega_2^0$ in Eqs. (\ref{eq:omega10})
and  (\ref{eq:omega20}) can be determined by using the conservation of 
angular momentum. Note that there is no external torque on the pulsar, 
the dynamics is affected solely due to the internal density fluctuations. 
Thus the angular momentum is conserved. (We are neglecting the possibility
of significant emission of any particles during the phase transition.) Prior 
to the phase transition, the pulsar rotates about the z-axis with angular 
velocity $\omega$ and the angular momentum has only z-component $L_z = L$. 
After completion of the phase transition at $t = 0$, the orientations of 
the new set of principal axes $x_0$, $y_0$ and $z_0$ of frame $S_0$ are 
changed relative to the original frame $S$ as shown in Fig. \ref{fig:fig1}. 
The new $y_0$-axis is chosen to lie in the y-z plane, making 
an infinitesimal small angle $\alpha$ (for small perturbations) with the y-axis. 
Note that for the unperturbed state, the principal axis corresponding to $I_3^0$ is 
unambiguously fixed. However, this is not true for the other two axes 
(since $I_1^0 = I_2^0$) lying in the x-y plane. We have the freedom of 
choosing one of them arbitrarily. Here we have exploited this freedom 
to choose the y-axis (by rotating the x-y plane) in such a way that $y_0$-axis 
lies in the y-z plane. With this choice, we can now write the unit 
vector along $y_0$ as $\hat{y_0} = \hat{y} + \alpha~\hat{z}$. Denoting
the polar angle and the azimuthal angle of $z_0$-axis as $\theta_0$ and $\phi_0$, 
respectively (these are the standard angles in spherical coordinates measured 
relative to S-frame), one can also write the unit vector along $z_0$ as $\hat{z_0} 
= \theta_0 \cos\phi_0~\hat{x} + \theta_0 \sin\phi_0~\hat{y} + \hat{z}$. 
Using orthogonality, the angle $\alpha$ and the
unit vector $\hat{x_0}$ can be fixed as $\alpha = - \theta_0 \sin\phi_0$
and $\hat{x_0} = \hat{x} - \theta_0 \cos\phi_0~\hat{z}$.
Note that by expressing the unit vectors ($\hat{x_0}, \hat{y_0}, \hat{z_0}$) in terms
of ($\hat{x}, \hat{y}, \hat{z}$) allows us to determine the rotational matrix $R_0$, which 
describes the orientations of the new set of principal axes relative to the old set. 
The matrix $R_0$ is parameterized by the angles $(\theta_0, \phi_0)$ and can be written as

\begin{equation}
R_0 = 
\begin{pmatrix} 
1 & 0 & -\theta_0 \cos\phi_0 \\
0 & 1 &  -\theta_0 \sin\phi_0 \\
\theta_0 \cos\phi_0 & \theta_0 \sin\phi_0 & 1 \label{eq:rot}
\end{pmatrix}
\end{equation}

We will see later (section \ref{section:sec4}) the role of $R_0$ in 
our numerical calculations. The initial angle $\theta_0$ and $\phi_0$ are determined by 
diagonalizing the perturbed MI matrix, and finding the eigen vectors corresponding to 
three eigen values.

With the above choice 
of orientations of the new set of principal axes, we now resolve the original 
angular momentum $L_z$ ($ = L$) along $x_0$, $y_0$ and $z_0$, respectively. 
The corresponding components can be written as

\begin{align}\label{eq:comp1}
L_{x_0} (t=0) &= I_1 \omega_1^0 = - L \theta_0 \cos \phi_0 \\
L_{y_0} (t=0) &= I_2 \omega_2^0 = - L \theta_0 \sin \phi_0 \\
L_{z_0} (t=0) &= I_3 \omega_3^0 = L.
\end{align}

Using the above set of equations, the angular frequencies 
Eq. (\ref{eq:omega3}, 
\ref{eq:omega10}, \ref{eq:omega20}) can be 
expressed in terms of $\theta_0$ and $\phi_0$ as

\begin{align}
\omega_1 (t)&=\dot\theta_1=-\omega \theta_0 [\cos \phi_0 \cos (\Omega t) 
- \frac{\sin\phi_0}{k} \sin(\Omega t)]\label{eq:soln1}\\
\omega_2 (t)&=\dot \theta_2 = -\omega \theta_0 [k \cos \phi_0 \sin (\Omega t) 
+ \sin\phi_0 \cos(\Omega t)]\label{eq:soln2} \\
\omega_3 (t)&=\dot \theta_3 =\omega.\label{eq:soln3}
\end{align}

In the above, we have taken an approximation, $L/I_1 \simeq 
L/I_3 \simeq \omega$. This is in view of Eq. (\ref{eq:soln3}) and the fact
that angle $\theta_0$ is very small for tiny density fluctuations,
as we will see below. The numerical 
algorithm for finding the set of solutions $\theta_i(t)$ ($i =1,2,3$), and their 
role in modulating pulse profiles will also be discussed therein. Now before 
presenting such numerical prescription, we will provide below a few 
estimates of various quantities relevant to the precession.

\subsection {The choice of parameters and estimates of various quantities 
characterizing the precession}

First, we estimate the precession frequency $\Omega = [(I_3 - 
I_1)(I_3 - I_2)/(I_1 I_2)]^{1/2} \omega$. As the perturbations 
$\delta I_{ij}$ are small, the new set of principal axes are expected 
to be very close to the original (unperturbed) axes. This is also observed 
numerically to be discussed in the next section. Thus the principal MI 
of the perturbed state can be written as, $I_{1,2} = I_0 (1-\eta + 
\epsilon_{1,2})$ and $I_3 = I_0 (1+\epsilon_3)$. Where, $\epsilon_i$ 
($i = 1,2,3$) are taken to be of order $\epsilon$ for which we will take two
sample values, $10^{-8}$ and $10^{-5}$. The precession frequency 
$\Omega$ can then be expressed in terms of $\eta$ and $\epsilon$ (a 
function of $\epsilon_i~$ ; $i = 1,2,3$) as

\begin{equation} \label{eq:prece} 
\Omega \simeq \frac{\eta+\epsilon}{1-\eta+\epsilon} \omega \simeq \eta ~\omega.
\end{equation} 

Where, as mentioned above we have assumed that $\epsilon,~\eta <<1$ and
$\epsilon << \eta$. Therefore, the precession frequency is
completely determined by the deformation parameter $\eta$ of the 
unperturbed pulsars. Thus, for a millisecond pulsar, 
for example, the time period of precession will be of order 1 sec,
if the deformation parameter is of order $10^{-3}$. We will discuss the 
implications of this further in section \ref{section:sec5}.

The amplitude $\omega_m$ of frequency oscillations $\omega_1(t)$, and the 
amplitude $\theta_m$ of precession angle $\theta_1(t)$ can be estimated from Eq. 
(\ref{eq:soln1}). Note that since $k = [I_1(I_3 - I_1)/
(I_2(I_3 - I_2))]^{1/2} \simeq 1 + \epsilon/2\eta$, the corresponding quantities associated with 
$\omega_2(t)$ will be of same order as for $\omega_1(t)$.
The relative angular shift $(\theta_0, \phi_0)$ of the principal axes (see Fig. \ref{fig:fig1}) 
are determined by diagonalizing the perturbed matrix $I_{ij}$, and finding the eigen vectors 
corresponding to three eigen values. These eigen-vectors will then correspond 
to the set of three principal axes. The identification of $z_0$-axis 
can be done by finding the direction cosines of the eigen-vector corresponding 
to the largest eigenvalue. Now, as mentioned earlier, the perturbed 
moment of inertia (MI) matrix elements were taken to be $I_{ij} = I^0_{ij} 
+ \delta I_{ij}$. Where $\delta I_{ij}$ is assumed to be Gaussian distributed 
with width $\sigma = \epsilon I_0$ ($I_0 \equiv I^0_3$). For an analytical 
estimate of $\theta_0$, let us define a quantity $\epsilon_{ij}$, which 
characterizes the relative perturbation of MI matrix element due to the density 
fluctuations as $\epsilon_{ij} = \delta I_{ij}/I_0$. As the width $\sigma$ of 
the perturbations is assumed to be of order $\epsilon I_0$, all the components 
of $\epsilon_{ij}$  ($i, j = 1,2,3$) are also expected to be of order 
$\epsilon$. For a simple analytical estimate, we use the approximation 
$\epsilon_{ij} = \epsilon/I_0$. With this, the diagonalizations of the 
perturbed matrix gives the result,

\begin{equation} \label{eq:theta} 
\cos \theta_0 = \left (1+ 2\left(\frac{\epsilon I_0}{I_3 - I_1-\epsilon I_0}\right)^2 
\right)^{-1/2}.
\end{equation}

Substituting $I_1 = I_2 \simeq I_0 (1-\eta + \epsilon)$, $I_3 
\simeq I_0 (1+\epsilon)$ and assuming $\epsilon << \eta$, we now get 
the angular shift of $z^\prime$-axis (to leading order in $\epsilon$) as  

\begin{equation} \label{eq:theta0} 
\theta_0 \simeq \sqrt{2}~\left({\epsilon \over \eta}\right).
\end{equation}

 Allowing for slightly more general $\epsilon_{ij}$ also gives
similar result. The numerical procedure for finding $\theta_0$ 
will be discussed later in section \ref{section:sec4}. It turns out that 
for a general random values of $\epsilon_{ij}$, our numerical results  
also approximately produce the above analytical estimate of $\theta_0$.

As $k = [I_1(I_3 - I_1)/
(I_2(I_3 - I_2))]^{1/2} \simeq 1 + \epsilon/2\eta$, we can now rewrite 
Eq.(\ref{eq:soln1}) and Eq.(\ref{eq:soln2}) as

\begin{align}
\omega_1 (t) &= - \omega \theta_0 [\cos (\Omega t+\phi_0) + 
\frac{\epsilon}{2\eta} \sin\phi_0 \sin (\Omega t)] \label{eq:amp1}\\
\omega_2 (t) &= - \omega \theta_0 [\sin (\Omega t+\phi_0) + 
\frac{\epsilon}{2\eta} \cos\phi_0 \sin (\Omega t)] \label{eq:amp2}.
\end{align}

The corresponding rotational angles can be written as

\begin{align}
\theta_1 (t) &= - \frac{\omega \theta_0}{\Omega} [\sin (\Omega t+\phi_0) - 
\frac{\epsilon}{2\eta} \sin\phi_0 \cos (\Omega t)] \label{eq:amp3}\\
\theta_2 (t) &=  \frac{\omega \theta_0}{\Omega} [\cos (\Omega t+\phi_0) + 
\frac{\epsilon}{2\eta} \cos\phi_0 \cos (\Omega t)] \label{eq:amp4}.
\end{align}

Since $\theta_0 \simeq \sqrt{2}(\epsilon/\eta)$, the second
terms in the above set of equations (Eq. (\ref{eq:amp1}) - Eq. (\ref{eq:amp4})) 
are of order $\sim (\epsilon/\eta)^2$.
The resulting amplitude $\omega_m$ of frequency oscillations $\omega_{1,2}(t)$ 
(Eq. (\ref{eq:amp1}) and Eq. (\ref{eq:amp2})), and the amplitude $\theta_m$ of 
precession angles $\theta_{1,2}$  ((Eq. (\ref{eq:amp3}) and Eq. (\ref{eq:amp4})) 
are thus given  by $\omega_m = \omega \theta_0 
\simeq \sqrt{2}(\epsilon/\eta) \omega$ and $\theta_m = 
(\omega/\Omega)\theta_0 \simeq \sqrt{2}~(\epsilon/\eta^2)$.
So for $\eta = 10^{-3}$, the oscillation amplitudes for $\theta_1$ and $\theta_2$ 
will be of order $10^{6}~\epsilon$. This, for example, results in 
approximately $1^\circ$ amplitude for $\epsilon = 10^{-8}$. The 
observational aspects of this significant result will be discussed 
in section 5. 

The effects of precession will be seen on the observed fluxes from 
the pulsars.  To estimate that, we assume the flux emission to be
conical in nature with the vertex of the cone at the centre of the
pulsar. The angular flux distribution for the 
pulsars is taken to be azimuthally symmetric about the centre of the 
emission region and is taken to be of Gaussian shape \citep{kris83} 
of width $w$, 
   
\begin{equation} \label{eq:flux1}
F(\alpha) = F_0~e^{-\frac{\alpha^2}{w^2}}.
\end{equation}

Where $\alpha$ is the angle of the radial vector of the emission point 
from the central axis of the cone.  If the perturbations result in the 
change of $\alpha$ by a small amount $\delta$, the corresponding change 
in flux becomes

\begin{equation} \label{eq:flux2}
F^\prime (\phi) = F_0~e^{-\frac{(\alpha + \delta)^2}{w^2}} 
\simeq F (\phi) \left(1 - 2\delta \frac{\alpha}{w^2}\right).
\end{equation}

Thus we see that the fractional change of flux will be of order $O(\delta)$, 
which is approximately equal to $\theta_m$, i.e., of order $\epsilon/\eta^2$.
Here, we recall our earlier discussion that the pulse modifications due to induced 
wobbling (from off-diagonal MI components) may be more easily observable, even if 
direct effects on the pulsar frequency arising from free energy difference between 
two thermodynamic phases remain suppressed.

\section{The algorithm for studying pulse modulations and the numerical 
results}
\label{section:sec4}

\begin{figure}
\centering
%\vspace{-1.0cm}
\includegraphics[width=0.6\linewidth]{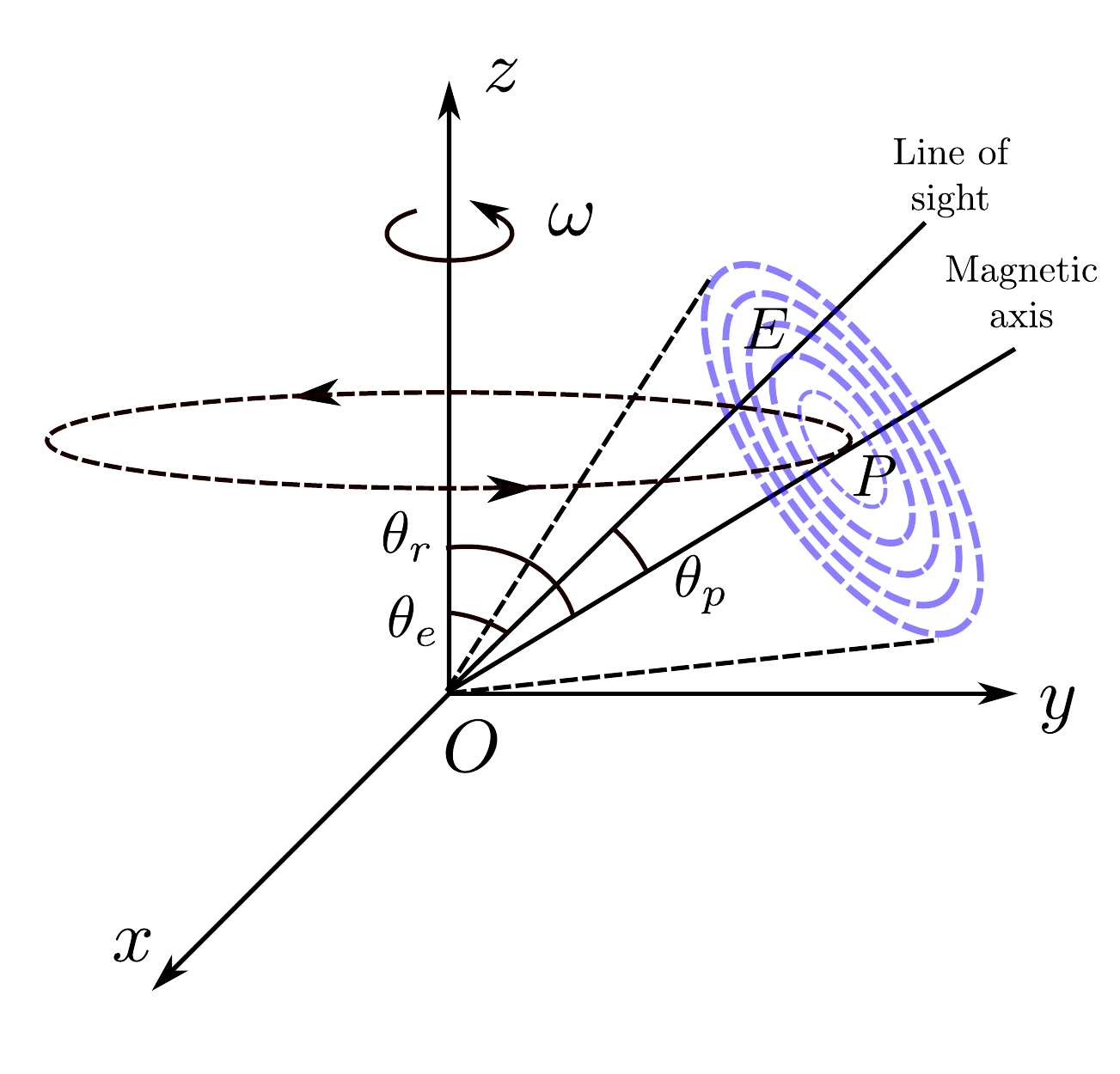}
%\vspace{-3.2in}
\caption{Figure shows the radiation emission cone of a pulsar. The magnetic 
axis (OP) and the line of sight (OE) pointing towards earth, make angle 
$\theta_r$ and $\theta_e$, respectively with the rotation axis. 
$\theta_p$ is the angle between OP and OE.}
\label{fig:fig2}
\end{figure}

We will describe here the numerical approach that was followed to study the 
effects on pulse profile due to the precession of a pulsar. First, we 
consider the profile for an unperturbed pulsar rotating freely about the 
z-axis with frequency $\omega$. We assume the standard conical shape 
geometry \citep{gil81, gil84} for the pulse emission region 
(Fig. \ref{fig:fig2}). The angles of magnetic axis and the line of sight 
pointing towards earth with the rotation axis are denoted by $\theta_r$ and 
$\theta_e$, respectively. Assume P$(R_r \sin\theta_r \cos\phi_r, R_r \sin\theta_r 
\sin\phi_r, R_r \cos\theta_r)$ and E$(R_e \sin \theta_e \cos\phi_e, 
R_e \sin\theta_e \sin\phi_e, R_e \cos\theta_e)$  to be the center of the 
pulse emission region, and the intersection point on the emission region 
by the direction vector towards earth, respectively.  Both the points are 
assumed to be on the surface of the pulsar, and $R_r \simeq R_e \equiv R$
is the radius (R) of the (almost spherical) pulsar.
%%%%%%%%  FIGURE-2 %%%%%%%%%
%%%%%%%%%%%%%%%%
The angle between $\vec {OE}$ and $\vec {OP}$ is denoted by 
$\theta_p$. Note that $\theta_p (t)$ changes with time as the pulse emission 
cone sweeps across the line of sight with rotation frequency $\dot \phi_r (t) 
= \omega$. The evolution of $\theta_p$ will be reflected in the observed intensity 
distribution of the pulses. For that, we take the intensity distribution of 
pulses as Gaussian \citep{kris83} of width $w$ (as in Eqn.(21)),

\begin{equation} \label{eq:profile}
I(\theta_p) = I_0~e^{-(\theta_p^2/w^2)}.
\end{equation}

Now, in the presence of density fluctuations, the above profile will be 
modulated due to the precession of pulsars. For understanding this, let us 
first set our notations and symbols for various quantities. Let $S(x,y,z)$ 
be the space-fixed frame (frame of the observer) with respect to which the 
pulse profile is supposed to be analyzed. $S_0(x_0,y_0,z_0)$, $S_1 
(x_1,y_1,z_1)$, ...are the instantaneous space-fixed frames, which coincide 
with the body-fixed frames at time $t = 0$ (immediately after the phase 
transition), $\Delta t, 2\Delta t, ...$ and so on. At an 
arbitrary time $t$, the orientations of body-frame axes are determined in terms
of the rotations $\theta_1 (t)$, $\theta_2 (t)$, and $\theta_3 (t)$ w.r.t. the 
corresponding instantaneous space-fixed frame, with corresponding 
angular frequencies $\omega_1 (t)$, $\omega_2 (t)$ and $\omega_3 (t)$, 
respectively. Assume an arbitrary fixed point $P^*$ on the surface of the 
(almost perfectly spherical) star, whose angular coordinates with respect to 
the space-fixed frame $S$ at time $t = 0$, are labeled by 
$(\theta^*, \phi^*)$. After the phase transition, at $t = 0$, the new 
principal axes become different, given by the body-fixed frame $S_0$, 
without any rotation occurring for the body. Hence, the location of this point 
$P^*$ w.r.t the body-fixed frame $S_0$ will always be given by 
$R_0 (\theta^*,\phi^*)$ at any time $t$. Here $R_0$
is the rotational matrix parameterized by the angles ($\theta_0, \phi_0$)
describing the orientations of $S_0$-frame relative S-frame (see Eq.(\ref{eq:rot})).
As the body rotates, the corresponding angular coordinates as seen by a space-fixed 
observer at an arbitrary time $t$ are denoted by $(\theta (t), \phi (t))$. 
Corresponding Cartesian coordinates will be represented as column vectors  while 
performing coordinate transformation through the operation of rotation 
matrix.  The matrices $R_x (\theta_1)$, $R_y (\theta_2)$ and $R_z 
(\theta_3)$ describe the rotations by angle $\theta_1$ about x-axis, 
$\theta_2$ about y-axis and $\theta_3$ about z-axis, respectively. The 
rotation matrices $R_0$, $R_1$, ..., respectively describe the orientations 
of \lq $S_0$-frame relative to $S$-frame\rq, \lq$S_1$-frame relative to 
$S_0$-frame\rq,...and so on. These matrices are in turn the products 
of matrices, $R_x$, $R_y$ and $R_z$. As the rotations are being considered
for infinitesimal time period $\Delta t$ w.r.t. the instantaneous space-fixed
frames, all the angles are infinitesimal, and  $R_x$, $R_y$ and $R_z$ 
commute with each other. 

We will now discuss below our algorithm using which the effect of 
precession on pulse profile is calculated. As we 
mentioned above, diagonalization of the perturbed MI matrix gives the new set of 
principal axes ($S_0$-frame in Fig. \ref{fig:fig1}). The orientations of axes of $S_0$ relative to 
those of $S$ is obtained through $R_0$ which is parameterized by the initial angles 
$\theta_0$ and $\phi_0$ of the $z_0$ axis.
We noted above that the coordinates of the radiating point $P^*$ in the body fixed frame 
at any time $t$ are fixed, given by $R_0 (\theta^*,\phi^*)$.  After this initial 
set up, following steps are performed to get the pulse profile for the perturbed 
pulsar.

Step - 1 ($t = \Delta t$) : $\theta_i (\Delta t)$, ($i = 1,2,3$) 
is obtained by integrating Eq. (\ref{eq:soln1}) - Eq. (\ref{eq:soln3}) for 
a time step $\Delta t$.  The matrix $R_1$ that describes the orientations 
of $S_1$-frame relative to $S_0$-frame is obtained through $R_1 = R_x 
(\theta_1)R_y (\theta_2)R_z (\theta_3)$. We then get the location of 
the point $P^*$ at time $ t = \Delta t$ as seen by the space-fixed
fixed observer (frame $S$) through the coordinate transformations, 
$[\theta (\Delta t), \phi (\Delta t)] = R_0^{-1} R_1^{-1} R_0 (\theta^*, \phi^*)$. 
Note that above prescription is valid for any arbitrary point $P^*$ on the 
surface of the star. For calculating $\theta_p$, the point $P^*$ is chosen as 
the center of the emission cone labeled as "P" in Fig. \ref{fig:fig2}. As the 
star rotates, the angular coordinates of this point change w.r.t. the space 
fixed frame $S$ leading to changing $\theta_p$.  Following the same procedure 
as one would do for the unperturbed pulsars, $\theta_p$ is calculated at time 
$t = \Delta t$ and hence, the intensity of the pulse $I (\theta_p)$ is 
obtained from Eq. (\ref{eq:profile}).

Step - 2 ($t = 2 \Delta t$) : Following the same prescription as 
above, $\theta_i (2 \Delta t)$, ($i = 1,2,3$) is obtained for the next time 
step $\Delta t$ (Note, the integration is performed from 
$\Delta t$ to 2$\Delta t$).  This allows to determine the matrix $R_2$, which relates 
$S_2$ with $S_1$ through $R_2 = R_x (\theta_1)R_y (\theta_2)R_z (\theta_3)$. 
The location of $(\theta^*, \phi^*)$ at time $2 \Delta t$ relative to $S$ is 
obtained through $[\theta (2 \Delta t), \phi (2 \Delta t)] = R_0^{-1} 
R_1^{-1} R_2^{-1}R_0 (\theta^*, \phi^*)$. Again, $I (\theta_p)$ is obtained 
at time $t = 2 \Delta t$ using Eq. (\ref{eq:profile}) . 

The above time steps are repeated for a sufficiently long time duration to 
observe the modulations of the pulse profile due to precession.
For clarification, here we should mention that the set of matrices 
($R_1, R_2,$...) represents the sequence of time evolution. However each of 
these rotation matrices itself consists of three rotation matrices, about x,y, 
and z axis, respectively. For example, the matrix $R_1$ is given by 
$R_1 \equiv (R_x({\theta_1}) R_y({\theta_2}) 
R_z({\theta_3}))$, and similarly for $R_2$ and so on. We take these three matrices 
$(R_x({\theta_1}) R_y({\theta_2})R_z({\theta_3}))$ to 
commute as they represent infinitesimal rotations for small time interval $\Delta t$. 
However, $(R_1, R_2,...)$ in sequence represent time integration of rotations. These 
are naturally time ordered and we do not assume their commutation.

It should also be noted that the calculation of the time evolution 
of any fixed point due to precession followed by coordinate transformation to S-frame
necessitates the appearance of $R_0$ matrix twice. The first $R_0$ matrix 
(which now involves two angles $\theta_0$ and $\phi_0$) gives the coordinates of the 
radiation point P* in the ($x_0,y_0,z_0$) frame. 
The radiation point $P^*$ is taken to have coordinates ($\theta^*,\phi^*$) 
in the original space-fixed frame (which is now given by axes ($x,y,z$).
Immediately after the phase transition, the point $P^*$ does not move, but
the choice of axes now becomes the body fixed frame ($x_0,y_0,z_0$). The
coordinates of $P^*$ in this body fixed frame are always given by $R_0(P*)$.
As the body rotates, at each time step, the location of this point $R_0(P*)$ 
in the body fixed frame has to be transformed to the  original space-fixed frame 
($x,y,z$). This gives the sequence of matrices $(R_0^{-1}) (R_1^{-1})....$
%%%%%%%%  FIGURE-3 %%%%%%%%%
\begin{figure}
\centering
\vspace{-0.5cm}
\includegraphics[width=0.8\linewidth]{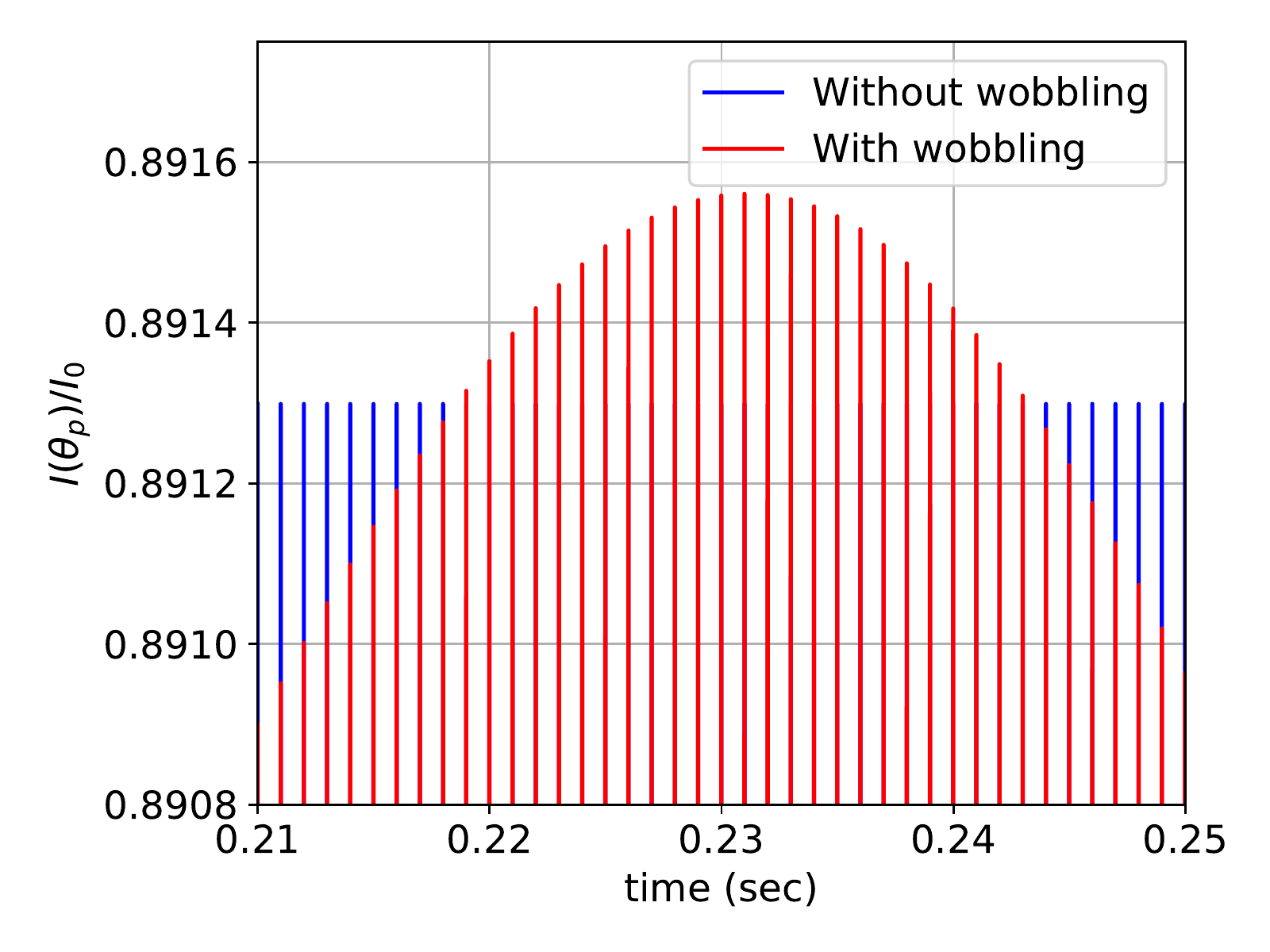}
%\vspace{-3.2in}
\caption{Evolution of normalized pulse intensity $I(\theta_p)/I_0$
with (red color) and without (blue color) modulation (induced by density 
fluctuations) for a millisecond pulsar, for the parameter set number 1 
as listed in Table 1.}
\label{fig:fig3}
\end{figure}
%%%%%%%%  FIGURE-4 %%%%%%%%%
\begin{figure}
\centering
\vspace{-0.5cm}
\includegraphics[width=1.0\linewidth]{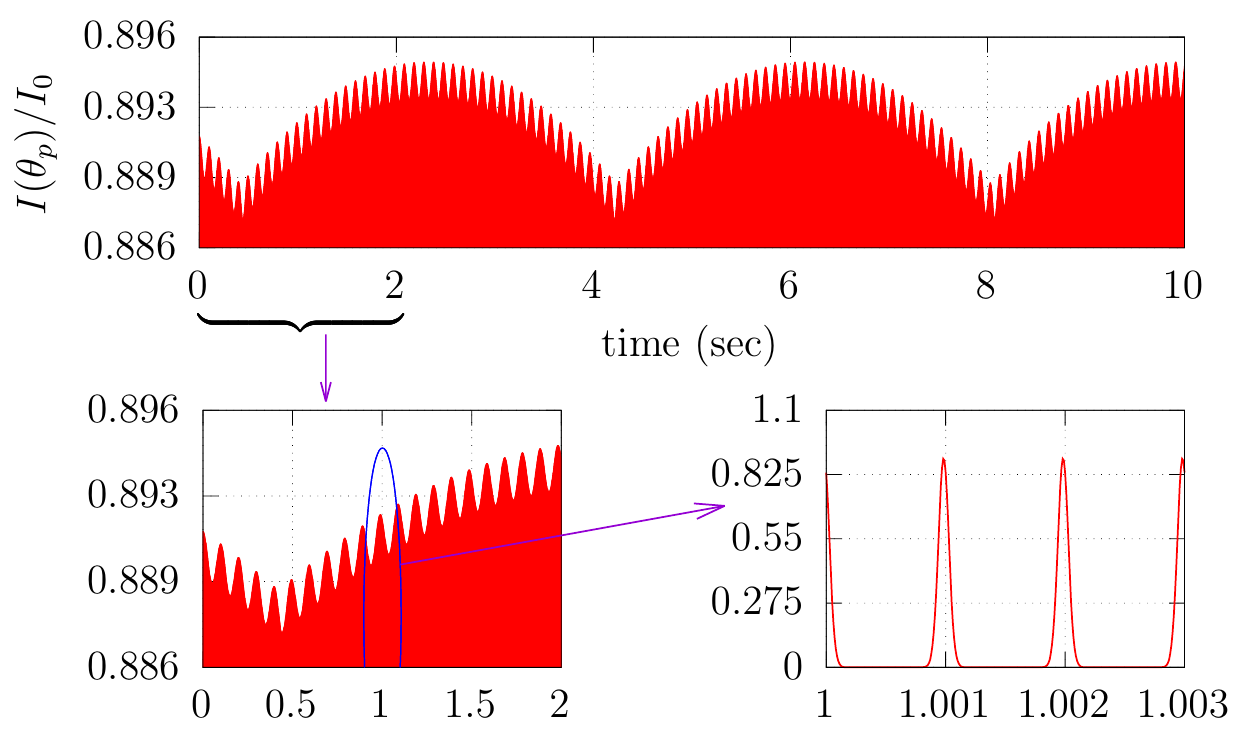}
%\vspace{-3.2in}
\caption{Time evolution of $I(\theta_p)/I_0$ in the presence of density
fluctuation induced modulation for the parameter set number 1
in Table 1.  Top plot shows the evolution of the top portion of the pulse
for a long time duration, clearly showing two different modulation
time scales. The plot interior is solid filled up due to crowding
of millisecond pulses. Bottom left  plot shows the same plot 
for a smaller time duration for a better resolution, which is further 
resolved (bottom right) to observe full profiles of a few individual 
millisecond pulses.}
\label{fig:fig4}
\end{figure}
%%%%%%%%  FIGURE-5 %%%%%%%%%F
\begin{figure}
\centering
\vspace{-0.5cm}
\includegraphics[width=1.0\linewidth]{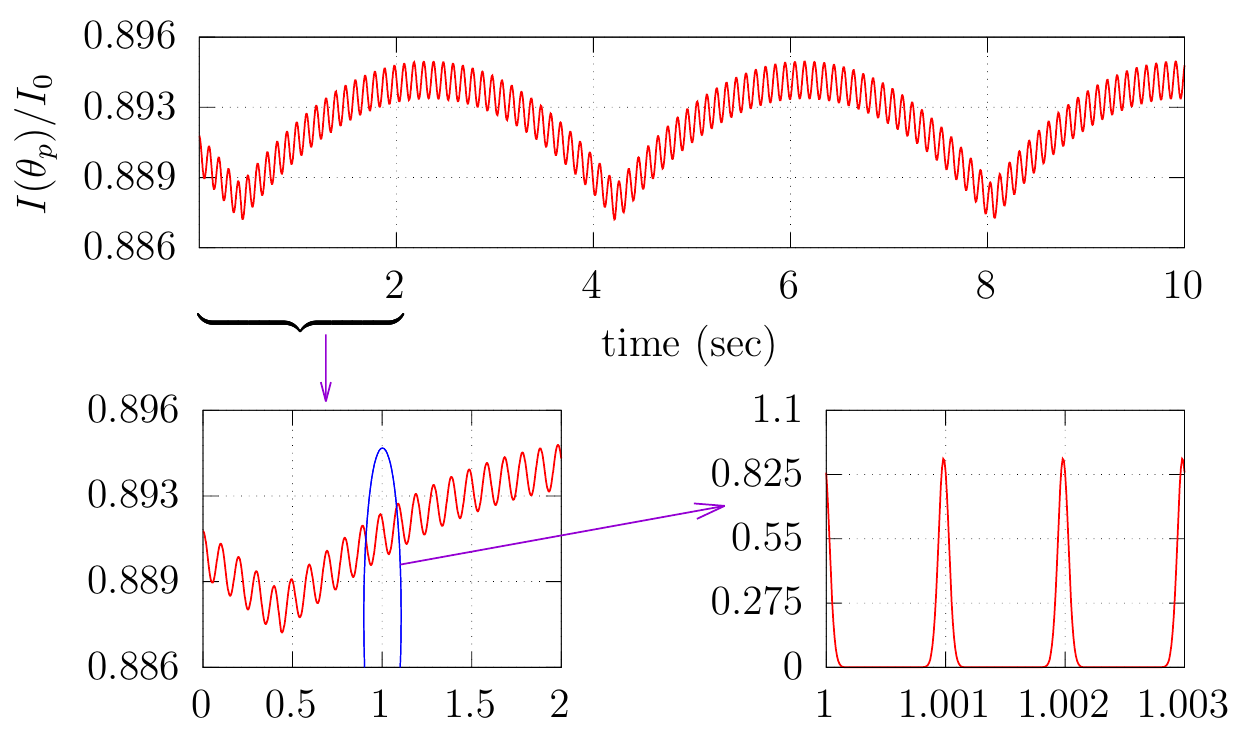}
%\vspace{-3.2in}
\caption{Same plots as in Fig. \ref{fig:fig4}, now only showing the top part of
the pulse profiles for clear visibility of the modulated pulse shape 
details.}
\label{fig:fig5}
\end{figure}

%%%%%%%%  FIGURE-6 %%%%%%%%%F
\begin{figure}
\centering
\vspace{-0.5cm}
\includegraphics[width=1.0\linewidth]{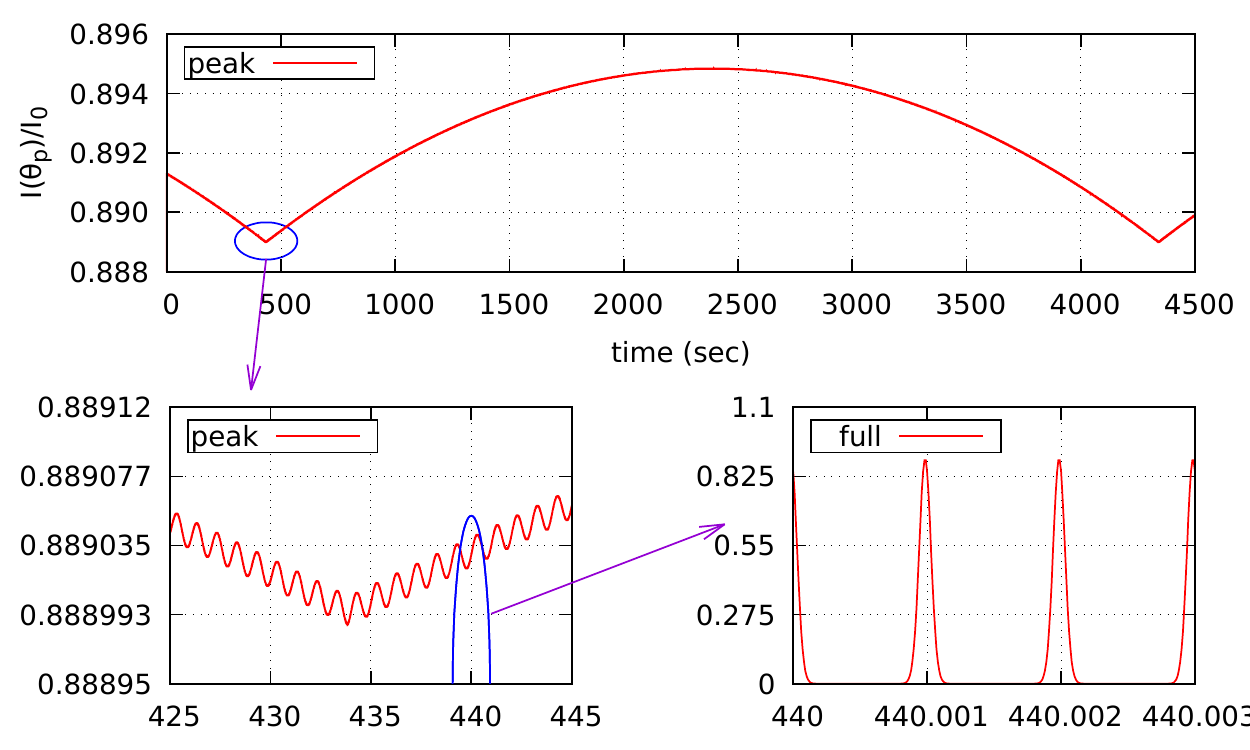}
%\vspace{-3.2in}
\caption{Same plots as in Fig. \ref{fig:fig5}, now for parameter set number 2 in Table 1.
Note, flux profile is perfectly smooth in the entire time domain. The 
apparent {\it kink}  which appears at time around 435 seconds 
in the top figure becomes smooth with an improved resolution as shown in 
the bottom left plot showing expanded plot in that region.} 
\label{fig:fig6}
\end{figure}
%%%%%%%%%%%%%%%%%%%
We will now present the results obtained using the above sequence of steps.
The parameters used in our calculations are listed in Table 1. A millisecond 
pulsar is chosen as a candidate for studying the effects of precession on 
pulse profiles. The angles of magnetic axis and the line of sight pointing
towards earth relative to the (unperturbed) rotation axis are taken as 
$\theta_r = 45^\circ$ and $\theta_e = 
40^\circ$, respectively (Fig. \ref{fig:fig2}). The initial (i.e., at $t=0$) 
azimuthals of the locations P and E  (Fig. \ref{fig:fig2}) are 
taken as $\phi_r = 45^\circ$ and $\phi_e = 40^\circ$, respectively.
Note, the choice for $\phi_r$ and $\phi_e$ at $t = 0$ has the 
same azimuthal separation $\Delta \phi$ as for the angular separation 
$\Delta \theta$ between $\theta_r$ and $\theta_e$. The same value of 
$\Delta \phi$ and $\Delta \theta$ is just an arbitrary choice and in 
principle, it could be anything. In fact, for other values of $\Delta \phi$, 
there will simply be a phase shift in the modulation. Now, as mentioned earlier, 
the change in MI components $\delta I_{ij}$ caused by the density fluctuations 
are assumed to be Gaussian \citep{kris83} with width $\sigma = 
\epsilon I_0$. The estimates in ref. \citep{bagchi15}, suggested that the 
values of $\epsilon$ may lie in the range $10^{-14}$ to $10^{-6}$.  Here, 
for a case study, we choose  two sample values of $\epsilon$ as $10^{-8}$ and 
$10^{-5}$.  We also use two values of the deformation parameter $\eta$ of the 
assumed oblate shape pulsar as $\eta = 10^{-3}$ and $\eta = 10^{-2}$.
(We again emphasize, we use these parameter values so that different
modulations have reasonable time period which can be seen in our numerical
simulations. The results are easily extended to much smaller values of
$\epsilon$ as well as $\eta$, which usually will lead to very long time
scales of modulations.) Note that the parameters $\eta$ and $\epsilon$ set 
the time scales for the expected flux modulations of the pulses due to the 
precession.  As we discussed above, this can be understood from the equations 
of motion (Eq. \ref{eq:soln1} and 
Eq. \ref{eq:soln2}) for $\omega_1$ (or $\omega_2$), which is
given  as $w_1 \sim \omega_m \cos (\Omega t)$. Thus, the time period 
$T_\Omega$ corresponding to the precession frequency $\Omega$ should set 
one of the time scales for the flux modulation. 
Now, we see from Eq. (\ref{eq:prece}) that (with our choice $\epsilon << 
\eta$), the precession frequency depends 
only on the pre-existing deformation parameter $\eta$ through $\Omega 
\simeq \eta \omega$. Thus, for a pulsar with time period $T_\omega = 
10^{-3}$ s, the characteristic time scale $T_\Omega  =  
T_\omega/\eta$ turns out to be about 0.1 second for $\eta = 10^{-2}$ 
and one second for $\eta = 10^{-3}$. As we will see below, this is precisely
what we see with our numerical results. 

Other than this modulation (let us call it as the {\it first} modulation), there 
will be another modulation time scale.  This is because $\omega_1$ and $\omega_2$ 
also  describe  periodic motions about $x$ and $y$ axis respectively. This should 
lead to another (say, the {\it second} modulation) of the millisecond pulses. The 
amplitude of $\omega_1$ oscillation is $\omega_m$ (Eq. (\ref{eq:amp1})), similar 
for $\omega_2$, as $k \simeq 1$.  With $\omega_m \sim (\epsilon/\eta)\omega
= (2\pi \epsilon/\eta)1000$ /sec., we expect the second modulation time scale
to be determined by the time scale $T_m \simeq 10^{-3}~(\eta/\epsilon) 
$ sec. The value of $T_m$ is of order of few seconds for 
$ (\eta, \epsilon) = (10^{-2}, 10^{-5}$), and a few hundred seconds for 
$(\eta, \epsilon) = (10^{-3}, 10^{-8})$. Note that this is the smallest
value of second modulation time scale expected because $\omega_m$ gives 
largest value of $\omega_1$ (and $\omega_2$), being the amplitude of
$\omega_1, \omega_2$ oscillations. As $\omega_1 (\omega_2)$ oscillates with
frequency $\Omega$, changing in magnitude from 0 to $\omega_m$, the final
time scale for the second modulation will be larger than the value $T_m$
estimated above. Further,  the complexity of precession of a rigid body in 
the presence of three rotations about the three 
axes can only be handled through 
the numerical simulations.  Our numerical results show that, indeed, there 
is a second modulation time scale of the pulses, and that its time scale
is larger by almost on order of magnitude than the value of $T_m$ estimated
above. 

For the initial Gaussian intensity distribution (Eq. (\ref{eq:profile})) we take the
angular width $w = 15^\circ$. The results from our numerical analysis are 
shown in Fig. \ref{fig:fig3} to Fig. \ref{fig:fig6}. For the time evolution of 
pulses, and calculations of $\theta_p(t)$ and $I(\theta_p)$, simulations are 
performed with time step $dt = 10^{-5}$ sec. This corresponds to total hundred 
time steps for each oscillation of a millisecond pulse. 
The time evolution of the normalized  flux $I(\theta_p)/I_0$ is shown in 
Fig. \ref{fig:fig3} for parameter set no.1  (see Table 1 for the choice of 
parameters). The red and blue colors, respectively, represent the evolution 
of the above quantities with and without precession. The effects of 
precession of pulsar are clearly imprinted in the modulation of
$I(\theta_p)/I_0$. The time scale $T_\Omega$ of about 0.1 second for this
parameter set for the first flux modulation is also visible in 
Fig. \ref{fig:fig3}. Due to small time duration of the plot, the second
modulation is not visible in this plot.

Fig. \ref{fig:fig4} shows the long time evolution of the pulse profile 
($I(\theta_p)/I_0$) in the presence of density fluctuation induced 
precession for parameter set no.1 in Table 1. This also shows the
second modulation, with typical time scale of roughly 5 seconds.
Recall, we estimated a time scale of about 1 sec. for this second 
modulation (for this parameter set). However, as discussed above,
this is because of using maximum value $\omega_m$ for $\omega_1$ and
$\omega_2$, and it is perfectly reasonable to get a larger time scale
for this second modulation. Top plot shows the evolution of the top 
portion of the pulse, clearly showing the two different modulation
time scales. The plot interior is solid filled up due to crowding
of millisecond pulses. Bottom left  plot shows the same plot 
for a smaller time duration for a better resolution, which is further 
resolved (bottom right) to observe full profiles of a few individual 
millisecond pulses. Fig. \ref{fig:fig5} shows the same plot as in Fig. 
\ref{fig:fig4}, but only showing the top of the pulse profiles for clear 
visibility of the modulated pulse shape details.

Simulations were also performed for a longer time duration for the  
parameters set number 2 in Table 1. The results are shown 
in Fig. \ref{fig:fig6}. Here we only show the top
of the pulse profile for clear visibility (as in Fig. \ref{fig:fig5} for parameter set 
no.1). The top figure shows the long time period modulation (the second
modulation) with time scale of few thousand seconds. We had estimated
time scale for the second modulation for this case to be few hundred
seconds. As discussed above for Fig. \ref{fig:fig5}, longer time period for
second modulation is reasonable to expect.  Note, flux profile 
is perfectly smooth in the entire time domain. The apparent {\it kink}  
which appears at time around 435 seconds 
becomes smooth with an improved resolution as shown in the bottom left 
plot showing expanded plot in that region. The characteristic time 
scale for the first flux modulation is clearly observed in this zoomed plot
and is of order one second, which agrees with our analytical estimate of 
$T_\Omega$ for this parameter set.  
%%%%%%%%%%%%  TABLE  %%%%%%%%%%
\begin{table}
\centering
\caption{Values of various parameters used in our calculations are listed
below. The parameter $\eta$ characterizes the deformation of the pulsar,
and $\epsilon$ is the fractional change of MI arising due to density 
fluctuations.  The angular width of the assumed Gaussian shape pulse 
profile is denoted by $w$.  $\theta_r$ and $\theta_e$ are the polar angles of 
the magnetic axis, and the line of sight pointing towards earth w.r.t.
the rotation axis, respectively.}
\label{tab:table}
\begin{tabular}{lcccccc} % four columns, alignment for each
\hline
Set number & $\eta$  & $\epsilon$ & $w$ & $\theta_r$ & $\theta_e$ \\
\hline
1 & $10^{-2}$ & $10^{-5}$ & $ 15^\circ $ & 
$ 45^\circ $ & $ 40^\circ $ \\
\hline
2 & $10^{-3}$ & $10^{-8}$ & $ 15^\circ $ 
& $ 45^\circ $ & $ 40^\circ $ \\
\hline
\hline
\end{tabular}
\end{table}
%%%%%%%%%%%%%%%%%%%%%%
\section{Various observational aspects of our results}
\label{section:sec5}
It is important to realize that the pulse modulations discussed
here resulting from wobbling of pulsar  due to density fluctuation will 
be necessarily transient. As the density fluctuations dissipate away, 
the pulsar will restore its original state of rotation (apart from
any effects of free energy changes to the new uniform phase as discussed
above). This should help in disentangling the
phase transition induced modulation from any other modulations 
present for the pulsar (e.g. due to any permanent non-uniformities in
the pulsar). One should look for transient changes in pulse profile
for any signal of phase transitions. We should mention that by no means
we imply that these two modulations are the only possible features of
the effects of phase transition induced density fluctuations on the
pulses. We have identified these two modulations as clear and distinct 
features. It will be interesting to find any other 
possible hidden patterns in these modified pulses. 
For example, \citet{jones01} (see also \citep{ira03,akgun06}) 
have studied the effects of precession on various aspects of
electromagnetic signal, such as arrival time residuals, pulse polarisation etc., 
arising from the electromagnetic spin-down torque. It will be interesting 
to quantify such effects in our model, where the pulsar precession is induced
by density perturbations.

In our present study, the time scale of the first modulation, 
with a shorter time scale should be possible to see in the pulsar data easily. 
The observation of longer time modulation may be much more difficult. It will 
depend on the entire time scale of the completion of the phase transition.
If the transition is completed (to a uniform new phase with no
density fluctuations present any more) in a relatively short period
(compared to the expected time scale of the second modulation), then
only small part of the modulation may be visible, and not the whole
cycle. This also brings in another important feature of these modulations.
As we have seen, the modulation time periods, as well as the amplitude
of modulation are proportional to the magnitude of density fluctuations
(characterized by $\epsilon$ here). The manner in which the density
fluctuations decay away after a phase transition depends crucially
on the nature of the transition. For example, during a first order
phase transition, density fluctuations typically decay away with
the time scale of coalescence of bubbles. For a continuous transition,
density fluctuations show scaling pattern with universal scaling
exponents. Very interesting possibilities arise when there are
topological defects produced in a phase transition. Coarsening
of domain wall defects, string defects etc. have been very well studied
in literature (see for example the review \citep{bran94}) and 
it is known that density fluctuations
due to these have specific scaling exponents, with energy density
scaling with time as $t^{-b}$ . Analytical calculations, as well as numerical
simulations show that $b = 1$ for string defects (see \citep{string1}
and \citep{string2}). 
Thus, by making detailed observation
of the changes in the pulse modulation amplitude as well as modulation
period, one should be able to identify the source of density fluctuation,
and hence the specific symmetry breaking pattern associated with the
phase transition occurring inside the pulsar. We again remind the reader 
that high density QCD transitions can lead to variety of topological defects. 
For example, transition to CFL phase, as well as the nucleonic superfluid transition 
can lead to string defects.

One important implication of our analysis points to a sort of
{\it memory effect} in the pulsar signal. As we mentioned, after all
density fluctuations fade away and uniform phase is achieved, the original
state of rotation will be restored, without any wobbling effects. So
no modulation of pulse profile will survive (assuming negligible effects
on pulsar frequency due to free energy difference between the two phases).
However, original state of rotation only means original angular velocities
about the original, unperturbed, principal axes. It does not mean
that one will get exactly same angular coordinates (say, of the radiation
emitting region) in later stages, as one would have obtained in the absence
of any phase transition. With intermediate change in the state of rotation
(angular velocities as well as new rotated principal axes), the location of
the angular coordinates at the complete end of phase transition will
depend on various details of the intermediate stage, along with the duration
and rate of restoration of the original state of rotation. In fact, in general
one will expect a shift in the angular position of the emitting region.
Thus, there should be a {\it residual} time shift in the pulsar signal
for any time after the end of the phase transition. Presence of any such
residual time shift in the pulsar signal can thus be attributed to an
earlier phase transition stage which could have been missed in direct
detection (say, by the modulation of pulses as discussed in this paper).
Of course, as discussed in \citep{jones01} a residual
time shift can have different origins as well.

\section{Conclusion}
\label{section:sec6}

 We have calculated  detailed modification of pulses from a pulsar
arising from the effects of phase transition induced density fluctuations 
on the pulsar moment of inertia. To represent a general
situation of such statistical density fluctuations, we have used a 
simple model where the initial moment of inertia tensor $I_{ij}^0$ of the
pulsar is assumed to get a random additional contribution $\delta I_{ij}$ for 
each of its component where  $\delta I_{ij}$ is taken to be Gaussian 
distributed with width $\sigma = \epsilon I_0$.  Using sample values of
$\epsilon$ and the pulsar deformation parameter $\eta$, 
we numerically calculate detailed pulse modifications 
by solving Euler's equations for the rotational dynamics of the pulsar.
We also give analytical estimates which can be used for arbitrary values of
$\epsilon$, though for very small values, the resulting pulse modifications
may be beyond current observations.  We show that there are very specific
patterns in the perturbed pulses which are observable in terms of
modulations of pulses over large time periods. In view of the fact that 
density fluctuations fade away eventually leading to a uniform phase in the 
interior of pulsar, the off-diagonal components of MI tensor also vanish 
eventually. Thus, the modification of pulses due to induced
wobbling (from the off-diagonal MI components) will also die away eventually.
This  allows one to distinguish these pulse modulations from the
effects of any wobbling originally present.
Though, even at such late stages when all density fluctuations
die away and no pulse modulation survives, one will expect, in general, 
a residual time shift of the pulses as restoration of original angular 
velocities does not imply restoration of the angular orientations as per the
original pulses. Such a residual time shift in a pulsar signal could thus
be attributed to an earlier phase transition.

 We emphasize that in representing the effect of density fluctuations on MI
tensor in terms of Gaussian distributed components $\delta I_{ij}$ with 
a single parameter $\epsilon$, we have ignored details of characteristic 
statistics of the density fluctuations which could 
differentiate between different types of phase transitions. Thus, the present
study is meant to focus on the gross features of the pulse modification, such
as the period and amplitude of pulse modification. We plan to 
consider detailed modification of the MI tensor depending on specific
phase transition, and see if observations of the perturbed signal are 
capable of distinguishing between different phase transitions. In the present
analysis also, some information of the details of phase transition is
contained in the manner in which density fluctuations decay away. 
In particular for a continuous transition, or for topological defect
induced density fluctuations, density fluctuations decay away with
specific universal exponents, which may be observable by making details
analysis of pulse modulations.

\section{Acknowledgments}
P. Bagchi would like to thank P. Zhuang (Physics department, 
Tsinghua University) and the Tsinghua University for the financial 
support during this work. We thank the anonymous reviewer for pointing
out an error and constructive suggestions on our earlier
manuscript.

\section{Data Availability}
No new data were generated or analysed in support of this research.

%%%%%%%%%%%%%%%%%%%%%%%%%%%%
\bibliographystyle{mnras}
\bibliography{pulsar} % if your bibtex file is called example.bib
\bsp	% typesetting comment
\label{lastpage}
\end{document}